\documentclass[10pt,letterpaper,journal,amsmath,amssymb]{IEEEtran}
\usepackage[cmex10]{amsmath}
\usepackage{amsfonts,amsthm,bm,amssymb}
\usepackage{array,cite,eqparbox,dcolumn}
\usepackage{graphicx}
\usepackage{subfig}
\usepackage{graphicx}% Include figure files
\usepackage{dcolumn}% Align table columns on decimal point
\usepackage{bm}% bold math
\usepackage{amsthm}
\newtheorem{definition}{Definition}%[section]
\newtheorem{theorem}{Theorem}%[section]
\newtheorem{lemma}{Lemma}%[section]
\graphicspath{{figures/pdf/}}
\DeclareGraphicsExtensions{.pdf}
\def\Re{\mathrm{Re}}
\def\Im{\mathrm{Im}}
\def\ket#1{| #1 \rangle}
\def\bra#1{\langle #1 |}
\def \fl#1{\lfloor #1 \rfloor}

\def\dim{\operatorname{dim}}

\def\rank{\operatorname{rank}}
\def\Span{\operatorname{span}}

\def\Ad{\operatorname{Ad}}
\def\G{\mathcal{G}}
\def\H{\mathcal{H}}

\def\L{\mathcal{L}}
\def\M{\mathcal{M}}

\def\R{\mathbb{R}}
\def\C{\mathbb{C}}
\def\SU{\mathbf{SU}}

\def\CC{\mathcal{C}}
\def\so{\mathfrak{so}}
\def\sp{\mathfrak{sp}}
\def\su{\mathfrak{su}}
\def\uu{\mathfrak{u}}
\def\u{\mathfrak{u}}

\begin{document}
\title{Symmetries on Spin Chains: Limited Controllability and Minimal Controls for Full Controllability}

\author{\IEEEauthorblockN{Xiaoting Wang\IEEEauthorrefmark{1}\IEEEauthorrefmark{4}
        Daniel Burgarth\IEEEauthorrefmark{2} and
	S G Schirmer\IEEEauthorrefmark{3}\IEEEauthorrefmark{4}\\
	\IEEEauthorblockA{
	\IEEEauthorrefmark{1}%
	          Department of Physics, 
		  University of Massachusetts at Boston,\\
		  100 Morrissey Blvd, Boston, MA 02125, USA\\
	\IEEEauthorrefmark{2}%
	          Physical Sciences Building, Penglais Campus,
        	  Aberystwyth University, \\ SY23 3BZ Aberystwyth, United Kingdom\\
        \IEEEauthorrefmark{3}%
                  College of Science (Physics), Swansea University,\\
            	  Singleton Park, Swansea, SA2 8PP, United Kingdom\\
	\IEEEauthorrefmark{4}%
                  Dept of Applied Mathematics \& Theoretical Physics,
                  University of Cambridge,\\
                  Wilberforce Road, Cambridge, CB3 0WA, United Kingdom\\
           Email: x.wang@damtp.cam.ac.uk, daniel@burgarth.de, s.schirmer@swan.ac.uk, sgs29@cam.ac.uk}}}

\date{\today}

\maketitle

\begin{abstract}
Symmetry is a fundamentally important concept in many branches of
 physics. In this work, we discuss two types of symmetries, external
 symmetry and internal symmetry, which appear frequently in controlled
 quantum spin chains and apply them to study various controllability
 problems.  For spin chains under single local end control when external
 symmetries exists, we can rigorously prove that the system is
 controllable in each of the invariant subspaces for both XXZ and XYZ
 chains, but not for XX or Ising chains.  Such results have direct
 applications in controlling antiferromagnetic Heisenberg chains when
 the dynamics is naturally confined in the largest excitation subspace.
 We also address the theoretically important question of minimal control
 resources to achieve full controllability over the entire spin chain
 space.  In the process we establish a systematic way of evaluating the
 dynamical Lie algebras and using known symmetries to help identify the
 dynamical Lie algebra.
\end{abstract}


%\pacs{02.20.Sv,03.67.-a,03.67.Hk,03.67.Lx,75.10.Pq}
% 02.20.Sv    Lie algebras of Lie groups
% 75.10.Pq    Spin chain models
%\maketitle

\section{Introduction}

Controllability is a fundamental concept in control theory in general,
and control of quantum systems in particular.  Any quantum system with a
sufficient number of controls becomes fully
controllable~\cite{Burgarth-Bose,Burgarth-10,Alastair,Burgarth}.
Therefore we are most interested in the problems where the system has
only a limited number of controls and often limited controllability
(e.g. subspace controllability).  Such limited controllability is
usually due to the existence of symmetries in the
Hamiltonians~\cite{Thomas,WS_symmetry}, which restrict the dynamical
Lie algebra (DLA) of the system~\cite{D'Alessandro-book}.
Previous literature on quantum controllability has mainly
focussed on the cases where either the system is fully controllable
(hence implying universal quantum computation~\cite{uni_qc}), or not
fully controllable but with a DLA that scales linearly or quadratically
with the system size.  In contrast, in this work, we would like to study
systems that are not fully controllable but with a DLA large enough for
universal quantum computation.  

There are simple criteria for controllability of bilinear systems in
terms of the Lie algebra rank condition~\cite{sussmann} similar to the Kalman rank
condition for linear systems.  However, verifying controllability for
quantum systems is challenging, not least because the dimension of the
DLA associated with a multi-partite quantum system usually grows
exponentially in the number of particles (such as qubits).  This
exponential scaling makes it impossible in most cases to verify the Lie
algebra rank condition numerically.  It is therefore important to have
general algebraic controllability results for certain classes of systems
such as spin chains with a few controls of a certain type.  In this
paper we derive such results for spin chains with isotropic and even
more anisotropic couplings.  Unlike spin chains with Ising-type coupling
such systems are usually controllable with very few controls acting on a
small subset of spins.  However, controllability is limited by the
existence of symmetries in the Hamiltonians.  For instance, it has been
shown using the propagation property that Heisenberg chains are fully
controllable given two non-commuting control acting on the first spin
\cite{Burgarth-Bose} but not when there is only single control acting on
the first spin.  In the latter case the controlled system has symmetries
and decomposes into invariant subspaces~\cite{WS_symmetry}, preventing full
controllability.  However, it has been observed that such systems appear
to be controllable on each invariant subspace, in particular, the
largest excitation subspace, whose dimension scales exponentially with
system size~\cite{WSBB}.

In this work we give a rigorous proof of this subspace controllability
result for XXZ chains and then apply similar techniques to discuss the
subspace controllability of a general XYZ spin chain.  This system is
interesting as it provides arguably the \emph{simplest model of a
universal quantum computer} one could imagine: a physical Hamiltonian
with a single control switch to do the computation.  We further show
that the same result does not hold for XX chains, where a single control
acting on the end spin in a chain can only give controllability on a
subspace whose dimension does not scale exponentially with system size.
In this case additional controls are needed, and we discuss the minimal
local control resources for full controllability in this context.

This paper is organized as follows: in Section~\ref{sec:model}, we
introduce different types of spin chains and define two fundamental
types of symmetry, external and internal, and their relations to
controllability.  In Section~\ref{sec:local}, we present a complete
discussion on spin chains under a single end control, and rigorously
prove that for both XXZ and XYZ chains, the system is controllable in
each invariant subspace, and that this result is robust if the control
field has a leakage on the neighboring spins. In
Section~\ref{sec:minimal}, we investigate the XXZ or XYZ chains for
various types of two controls and we find the minimal control resources
for full controllability on the entire Hilbert space.  In
Section~\ref{sec:xx}, we study the dynamical Lie algebra for an XX chain
subject to a single end control and investigate the controllability for
an XX chain subject to two and three controls.

\section{Model and Basics}
\label{sec:model}

For a quantum system composed of $N$ spins, we denote the standard Pauli
operators by $X,Y,Z$ and the local operator $R$ acting on the $k$-th
spin by $R_k$, i.e., $R_k=I \cdots I R I\cdots I$, where $I$ is the
identity on a single spin.

\textbf{System Hamiltonian:}  We consider a spin network composed of $N$
spin-$\frac{1}{2}$ particles with spin-spin interaction characterized by the
following Hamiltonian
\begin{equation}
  \label{eqn:H0}
   H_0 = \sum_{(m,n)}
         a_{mn} X_m X_n + b_{mn} Y_m Y_n + c_{mn} Z_m Z_n
\end{equation}
with the special cases $a_{mn}=b_{mn}=c_{mn}$ corresponding to isotropic
Heisenberg coupling, $a_{mn}=b_{mn}$ to XXZ-coupling, $a_{mn}=b_{mn}$
and $c_{mn}=0$ to XX-coupling, and $a_{mn}=b_{mn}=0$ to Ising coupling.
For XXZ-networks it is convenient to set $\gamma_{mn}=a_{mn}=b_{mn}$ and
$c_{mn}=\kappa\gamma_{mn}$. We also require all couplings $(m,n)$ in
(\ref{eqn:H0}) form a connected graph.

The constants $a_{mn},b_{mn},c_{mn}$ determine the coupling strengths
between nodes $m$ and $n$ in the network.  Special cases of interest are
chains with \textit{nearest-neighbor} coupling, corresponding, e.g., to
linear qubit registers in quantum information processing, for which
$a_{mn}=b_{mn}=c_{mn}=0$ except when $m=n\pm 1$.  A network is
\textit{uniform} if all non-zero couplings are equal, i.e., $a_{mn} \in
\{0,a\}$, $b_{mn}\in \{0,b\}$ and $c_{mn}\in \{0,c\}$.  Every spin
network has an associated simple \textit{graph representation} with
vertices $\{1,\dots,N\}$ determined by the spins and edges by non-zeros
couplings, i.e., there is an edge connecting nodes $m$ and $n$ exactly
if $\gamma_{mn}\neq 0$.

\textbf{Controllability:} The controlled quantum dynamics we are
interested in is characterized by the following Schr\"odinger equation:
\begin{equation}
\label{eqn:dynamics}
  \dot\rho = -\frac{i}{\hbar} \left[H_0+\sum_{j=1}^m f_j(t)H_j,\rho\right].
\end{equation}
where $H_0$ is the system Hamiltonian in (\ref{eqn:H0}) and $H_j$,
$j=1,2,\cdots, m$ is a series of control Hamiltonians with time-varying
amplitudes $f_j(t)$. We define the system to be controllable if the
dynamical Lie algebra $\L$ generated by $iH_j$, $j=0,1,\cdots, m$ is
equal to the largest Lie algebra $\u(2^N)$ or $\su(2^N)$. The definition
of controllability is very intuitive: it can be shown that if the system
is controllable, then any unitary process $U\in \SU(2^N)$ can be
generated from (\ref{eqn:dynamics}) under certain control sequence in
finite time; if $\L \subsetneq \su(2^N)$, then there exists some unitary
gate $U\in \SU(2^N)$ that can never be generated under
(\ref{eqn:dynamics})~\cite{D'Alessandro-book}.  The concepts of
controllability and dynamical Lie algebra are very important for both
theory and control applications, as they characterize the reachable set
of the control dynamics and have answered the question whether a given
control task can be achieved or not. However, calculating the dynamical
Lie algebra can become extremely difficult or even intractable as $N$
increases. Therefore, we hope to use other properties of the
Hamiltonians to infer information about controllability, and symmetry
does play such a role.

\textbf{Symmetries:} We consider two types of Hamiltonian symmetries:
external symmetry and internal symmetry~\cite{Thomas}.

\begin{definition}
Let $H_j$, $j=0,1,\cdots, m$, be a set of Hamiltonians for a given
quantum system.  If there exists a Hermitian operator $S$ such that
$[H_j,S]=0$ for all $j$ then $S$ is called an \textbf{external symmetry}
for the Hamiltonians; assuming $H_j$ are trace-zero, if there exists a
symmetric or antisymmetric operator $S$ such that $H_j^T S + S H_j = 0$
for all $j$, where $H_j^T$ is the transpose of $H_j$ then $S$ is called
an \textbf{internal symmetry}.
\end{definition}

From the definition, external symmetry implies that all $H_j$ can be
simultaneously diagonalized, while internal symmetry implies that the
dynamical Lie algebra $\L$ generated by $H_j$ is a subalgebra of the
orthogonal algebra $\so(2^N)$ or symplectic algebra
$\sp(2^N)$~\cite{Jacobson}.  In both symmetry cases, $\L$ is strictly
smaller than $\su(2^N)$ and the system is not controllable.  It is
useful to investigate which operators can be the external symmetries.

\textbf{Example~1.} For the system Hamiltonian (\ref{eqn:H0}), a simple
class of symmetry operators is of the form $S=A_1A_2\ldots A_N$, where
$A_k$ is a local operator on the $k$-th spin, i.e., $A_k= \alpha_1 X +
\alpha_2 Y + \alpha_3 Z+\alpha_4 I$.  $[H_0,S]=0$ then requires $ [X_m
X_n,A_mA_n]=0$ for any connected link $(m,n)$ in (\ref{eqn:H0}), which
shows that the nontrivial external symmetry operators are $X_1X_2\cdots
X_N$, $Y_1Y_2\cdots Y_N$ and $Z_1Z_2\cdots Z_N$, which are often known
as the \emph{parity symmetry}.  Hence, if the control Hamiltonians only
contain local Pauli operators in one direction, such as $Z$ direction,
with $H_j=F_j(Z_1,\ldots, Z_N)$, then $S_p=Z_1Z_2\cdots Z_N$ is the
corresponding parity symmetry and all Hamiltonians are invariant in each
of the two eigenspaces of $S_p$ with parity $+1$ and $-1$.

\begin{figure}
\begin{tabular}{c}
\centerline{\includegraphics[width=0.7\columnwidth]{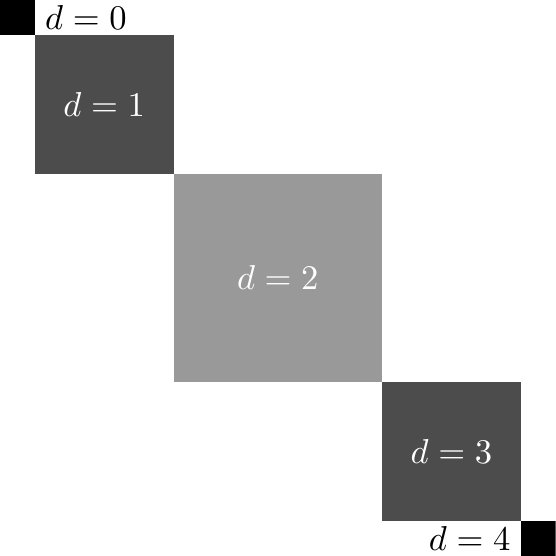}}
%\centerline{\includegraphics[width=5cm]{square-12}}
\end{tabular}
\caption{For $N=4$, the XXZ network Hamiltonian $H_0$ and local control
$H_1=Z_k$ are simultaneously block-diagonalized in 5 excitation
subspaces, $\H_d$, $d=0,\ldots,4$.}  \label{fig_excitation_sub}
\end{figure}

\textbf{Example~2.} If the system Hamiltonian (\ref{eqn:H0}) is of XXZ
type then defining $S_e=\sum_j^N(Z_j+I)/2$, we have $[H_0,S_e]=0$.
Physically, $S_e$ represent the total number of excitations, and has
$N+1$ distinct eigenvalues, ranging from $n=0$ to $n=N$, corresponding
to different numbers of excitations in the network.  If the control
Hamiltonians only contain $Z$ operators, i.e., $H_j=F_j(Z_1,\ldots,
Z_N)$ then $S_e$ defines an external symmetry, called the
\emph{excitation symmetry}, and all Hamiltonians are block-diagonalized
on the $N+1$ invariant subspaces, as illustrated in
Fig.~\ref{fig_excitation_sub}.

\begin{figure}
\begin{tabular}{c}
%\centerline{\includegraphics{figures/pdf/square-1.pdf}}\\
%\centerline{\includegraphics{figures/pdf/square-2.pdf}}\\
%\centerline{\includegraphics{figures/pdf/square-3.pdf}}\\
%\centerline{\includegraphics{figures/pdf/square-4.pdf}}
\centerline{\includegraphics{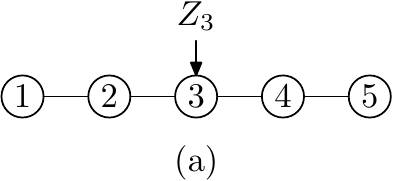}}\\
\centerline{\includegraphics{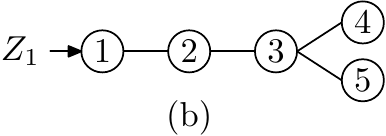}}\\
\centerline{\includegraphics{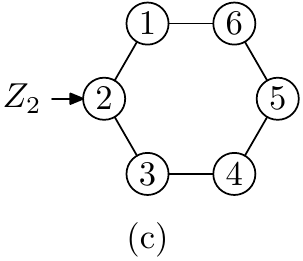}}\\
\centerline{\includegraphics{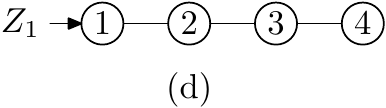}}
\end{tabular}
\caption{Different configurations \cite{Thomas} of controlled spin
networks with a local control $H_1=Z_k$ on the $k$-th spin indicated by
the arrows (the indexed circles represent different spins, with edges as
their couplings): (a),(b),(c) have permutation symmetry, (d) does not.}
\label{fig:permutation}
\end{figure}

\textbf{Example~3.} A non-identity element $\Pi$ in the permutation
group defines a \text{permutation symmetry} of the spin network if all
Hamiltonians are invariant under the permutation $\Pi$ of the spin
indices~\cite{WS_symmetry}.  In particular, for a single-local-control
problem such as $H_1=Z_k$ on the $k$-th spin, permutation symmetry means
that the index $k$ much be fixed under the permutation $\Pi$
(Fig.~\ref{fig:permutation}). In fact, $\Pi$ induces a symmetry operator
$S_{\Pi}$ which commutes with both the system and the control
Hamiltonians and hence defines an external symmetry.  Moreover, since
$S_{\Pi}$ commute with $S_e$ as defined in Example~2, $S_{\Pi}$ also
induces external symmetries on each excitation subspace of $S_e$, i.e.,
all Hamiltonians can be further block-diagonalized in the excitation
subspaces.

Having found all external symmetry operators of the Hamiltonians, the
entire Hilbert space can be decomposed into $\H=\oplus_{d=1}^D \H_d$,
where quantum dynamics is invariant on each $\H_d$, which cannot be
further decomposed. The associated dynamical Lie algebra $L$ must be a
subalgebra of $\oplus_{d=1}^D \uu(\dim \H_d)$.  Although the system is
not controllable on the entire space, it may still be controllable on
each $\H_d$.  In the following, we show that this is indeed true for
single local control $H_1=Z_1$ on the end spin of a XXZ chain, with the
symmetry operator $S_e$ as the total excitations.

\section{Single Local End Control}
\label{sec:local}

One of the simple but important configurations of a spin network is a
spin chain, which is the main subject of the paper.  We first consider a
spin chain with a single local control at the end of the chain. Without
loss of generality, we assume the control field is in Z direction.  The
corresponding controllability result depends on whether the spin-spin
interaction on the other two directions are equal or not, i.e., whether
the spin chain is of (1) XXZ type or (2) anisotropic XYZ type.

\subsection{XXZ Chain} 

For an XXZ chain with spin number $N$, under the end control in Z
direction, the system and the control Hamiltonians are written as:
\begin{subequations}
\label{eqn:xxz}
\begin{align}
 H_0&=\sum_{j}^N \lambda_j(X_jX_{j+1}+Y_jY_{j+1}+\kappa_j Z_jZ_{j+1})\\
 H_1&=Z_1
\end{align}
\end{subequations}
As discussed in previous section, the excitation operator $S_e=\sum_j^N
(Z_j+I)/2$ is an external symmetry, and the entire Hilbert space is
decomposed into $\H=\oplus_{k=0}^N \H_k$ with $\H_k$ as the invariant
subspace with $k$ excitations, i.e., it is generated by the
computational basis vectors with $k$ number of $1$'s, where the two
single-spin basis vectors are denoted as $\ket 0$ and $\ket 1$.  Hence
$\dim(\H_k)=\binom{N}{k}\equiv d_{N,k}$.  For example, for $N=4$, $\H_2$
is expanded by $\ket{0011}$, $\ket{0101}$, $\ket{0110}$, $\ket{1001}$,
$\ket{1010}$ and $ \ket{1100}$ with $\dim(\H_2)=d_{4,2}=6$.  Due to
$S_e$, the controlled system (\ref{eqn:xxz}) is not fully controllable
on the whole space, but it is controllable on each $\H_k$.  As an
application, for $\lambda_j<0$ when $H_0$ represents an
antiferromagnetic chain, and we can easily prepare the system into the
ground state $\psi_0$, which is in the largest excitation subspace
$\H_{\fl{N/2}}$ at $t=t_0$.  Then, by applying a single control $Z_1$
with amplitude $f(t)$ derived from optimization, we can generate the
total Hamiltonian $H=H_0+f(t)Z_1$ to drive the system into an arbitrary
target state in $\H_{\fl{N/2}}$ at a later time $t=t_F$. In particular,
we can generate perfect entangled pairs between the two end spins of the
chain, which is an important quantum resource for many applications such
as quantum communication or measurement-based quantum
computing~\cite{WSBB}.

Next we rigorously prove that under the control dynamics with the
Hamiltonians in (\ref{eqn:xxz}), the system is controllable in each
$\H_k$, and particularly in $\H_{\fl{N/2}}$.  By definition of
controllability, it is sufficient to show that $H_0$ and $H_1$ generate
$\uu(d_{N,k})$ on each $\H_k$.  Since $\H=\oplus_{k=0}^N \H_k$, the
associated dynamical Lie algebra $\L \subset \L_T=\oplus_{k=0}^{N}
\uu(d_{N,k})$. The idea of the proof is to determine all independent
operators generated in $\L$ and then evaluate $\dim(\L)$ in order to
identify $\L$.

Since a Lie algebra is also a real vector space, we can drop some
factors in the calculation and use linear combinations. We denote such
(trivial) steps in the derivation by $\rightarrow$. First of all, we
derive the following commutation relations:
\begin{align*}
[Z_1,H_0]&\rightarrow X_1Y_2-Y_1X_2\\
[Z_1,X_1Y_2-Y_1X_2]&\rightarrow X_1X_2+Y_1Y_2\\
[X_1X_2+Y_1Y_2,X_1Y_2-Y_1X_2]&\rightarrow Z_2-Z_1 \rightarrow Z_2\\
\cdots &\cdots
\end{align*}

Continuing this process, we can generate all $Z_j$, $X_jX_k+Y_jY_k$ and
$X_jY_k-Y_jX_k$ (with details in appendix \ref{A}). For brevity purposes
we will only focus on the $XX+YY$ terms and not write down the $XY-YX$
terms explicitly, since one operator can always be generated from the
other.  An operator is called a $k$-body operator if it contains $k$
nontrivial factors, i.e., those comprised of $X$, $Y$ or $Z$ Pauli
operators.  For example, $Z_1Z_3$ is a 2-body operator, while
$(Z_1-Z_2)Z_3Z_4$ is a 3-body operator. Denoting $M_k$ as the set of all
$k$-body operators in $\L$, we list all elements in $M_k$ and evaluate
$\rank{M_k}$:

(1) $M_1=\{Z_k\}$;

(2) $M_2=\{X_jX_k+Y_jY_k, \, Z_jZ_k\}$;

(3) $M_3=\{(X_jX_k+Y_jY_k)Z_m,\,(Z_j-Z_k)Z_mZ_n\}$;

(4) $M_4=\{(X_jX_k+Y_jY_k)Z_mZ_n, \\\null\qquad (X_jX_k+Y_jY_k)(X_mX_n+Y_mY_n), \,(Z_j-Z_k)Z_lZ_mZ_n\}$.

(5) $M_5=\{(X_{m_1}X_{m_2}+Y_{m_1}Y_{m_2})Z_{m_3}Z_{m_4}Z_{m_5},\\\null\qquad
    (X_{m_1}X_{m_2}+Y_{m_1}Y_{m_2})(X_{m_3}X_{m_4}+Y_{m_3}Y_{m_4})Z_{m_5},\\\null\qquad
    (Z_{m_1}-Z_{m_2})Z_{m_3}Z_{m_4}Z_{m_5}Z_{m_6}\}$;

$\cdots \cdots \cdots$

($\ell$) $M_\ell$: when $\ell$ is even, we can generate:
\begin{align*}
&(X_{m_1}X_{m_2}+Y_{m_1}Y_{m_2})Z_{m_3}\cdots Z_{m_\ell}\\
&(X_{m_1}X_{m_2}+Y_{m_1}Y_{m_2})(X_{m_3}X_{m_4}+Y_{m_3}Y_{m_4})Z_{m_5}\cdots
Z_{m_\ell}\\
& \cdots\cdots\cdots\\
&(X_{m_1}X_{m_2}+Y_{m_1}Y_{m_2})\cdots (X_{m_\ell-1}X_{m_\ell}+Y_{m_\ell-1}Y_{m_\ell})\\
&(Z_{m_1}-Z_{m_2})Z_{m_3}\cdots Z_{m_\ell};
\end{align*}
when $\ell$ is odd, we can generate:
\begin{align*}
&(X_{m_1}X_{m_2}+Y_{m_1}Y_{m_2})Z_{m_3}\cdots Z_{m_\ell}\\
&(X_{m_1}X_{m_2}+Y_{m_1}Y_{m_2})(X_{m_3}X_{m_4}+Y_{m_3}Y_{m_4})Z_{m_5}\cdots
Z_{m_\ell}\\
& \cdots\cdots\cdots\\
&(X_{m_1}X_{m_2}+Y_{m_1}Y_{m_2})\cdots
(X_{m_\ell-2}X_{m_\ell-1}+Y_{m_\ell-2}Y_{m_\ell-1})Z_{m_\ell}\\
&(Z_{m_1}-Z_{m_2})Z_{m_3}\cdots Z_{m_\ell}
\end{align*}

Next, in order to get $\rank(M_{\ell})$, we first evaluate the number of the
operators in the form
\begin{align*}
(X_{m_1}X_{m_2}+Y_{m_1}Y_{m_2})\cdots
(X_{m_{2p-1}}X_{m_{2p}}+Y_{m_{2p-1}}Y_{m_{2p}}),
\end{align*}
which contains $p$ pairs of $(XX+YY)$ or $(XY-YX)$ operators. We will
call them p-pair operators. For a given $N$ and $p$ with $N\ge 2p>0$, we
denote the set of p-pair operators as $E_{N,p}$.  For example, for
$N=2p=6$, $(X_1X_2+Y_1Y_2)(X_3Y_4-Y_3X_4)(X_5X_6+Y_5Y_6)$ is a 3-pair
operator in $E_{6,3}$. Then the size of the set $E_{N,p}$ is obtained by
simple combinatorics as
\begin{align*}
\frac{2^p}{p!}\binom{N}{2}\binom{N-2}{2}\cdots\binom{N-2(p-1)}{2}=p!\binom{N}{p}\binom{N-p}{p}.
\end{align*}

However, not all of the elements in $E_{N,p}$ are linearly
independent. For example, for $N=4$ and $p=2$, we find
\begin{align*}
&(X_1X_2+Y_1Y_2)(X_3X_4+Y_3Y_4)-(X_1X_3+Y_1Y_3)\\
&(X_2X_4+Y_2Y_4)=
(X_1Y_4-Y_1X_4)(X_2Y_3-Y_2X_3),\\
&(X_1Y_2-Y_1X_2)(X_3Y_4-Y_3X_4)-(X_1Y_3-Y_1X_3)\\
&(X_2Y_4-Y_2X_4)=
(X_1Y_4-Y_1X_4)(X_2Y_3-Y_2X_3),\\
&(X_1X_2+Y_1Y_2)(X_3Y_4-Y_3X_4)-(X_1X_3+Y_1Y_3)\\
&(X_2Y_4-Y_2X_4)=
(X_1X_4+Y_1Y_4)(X_2Y_3-Y_2X_3)
\end{align*}

Similarly we can write down the other dependence relations. Altogether there are only $1/2!$ of
all $2$-pair operators that are linearly independent.  In general, we
will prove that only $1/p!$ of all $p$-pair operators are linearly independent, and
\begin{align}
\label{eqn:Enp}
\rank(E_{N,p})=\binom{N}{p}\binom{N-p}{p}
\end{align}
However, directly proving (\ref{eqn:Enp}) is very difficult as the
linear dependence relations can become very complicated for large $N$
and $p$.  Fortunately, we can convert this problem to evaluating the rank
of a set of polynomials on complex field $\C$ (with details in appendix
\ref{B}).  Therefore, for $\ell>2$,
\begin{align*}
 \rank(M_\ell)=\sum_{p=1}^{\fl{\ell/2}} \rank(E_{N,p})\binom{N-2p}{\ell-2p}+\binom{N}{\ell}-1.
\end{align*}
In the above, the extra combinatorial factor arises from different
choices for putting the $Z$ terms.  After some simplification, for both
$N=2m$ and $N=2m+1$,
\begin{align*}
\dim{\L}= &\sum_{\ell=1}^m \rank(M_\ell)\\ 
        = &\rank(E_{N,1}) 2^{N-2}+\cdots+\rank(E_{N,m})2^{N-2m}\\
          & +2^N-N+1\\
         =&\sum_{p=0}^{\fl {N/2}} \frac{N!2^{N-2p}}{p!^2(N-2p)!}-N+1=\binom{2N}{N}-N+1
\end{align*}
where the last equation is shown in Lemma~\ref{combina} in the appendix.

As discussed earlier, $\L \subset \L_T\equiv \oplus_{k=0}^{N}
\uu(d_{N,k})$, with
\begin{align*}
 \dim(\L_T) = \sum_{i=0}^{N}\dim(\uu(d_{N,k}))
            = \sum_{i=0}^{N}{\binom{N}{k}}^2
            = \binom{2N}{N}.
\end{align*}
All $k$-body $Z$-type operators, $k=1,\cdots,N$, generate a Cartan
subalgebra $\CC$ in $\L_T$, with $\dim(\CC)=2^N$. Notice that since we
can only generate coupled $k$-body $Z$-type operators, such as
$(Z_{m_1}-Z_{m_2})Z_{m_3}\cdots Z_{m_k}$ in $\L$, the rank of all
$Z$-type operators in $\L$ is $2^N-N+1$, i.e., there are $N-1$
independent $Z$-type operators not included in $\L$, but included in
$\L_T$. We hence have:
\begin{align*}
\dim(\L)\le \dim(\L_T)-N+1=\binom{2N}{N}-N+1=\dim(\L)
\end{align*}
It means that $\dim(\L)$ achieves the allowed maximal value, which is
true only when $\L$ is isomorphic to $\uu(d_{N,k})$ on each
$\H_k$. Hence, we have proved the following theorem:

\begin{theorem}
For an XXZ chain of length $N$ with a single local control on the end spin in Z
direction, the system is controllable on each of the $N+1$ invariant excitation
subspaces. 
\end{theorem}

In particular, this theorem holds for anti-ferromagnetic Heisenberg
chains, which rigorously justifies the numerical findings
in~\cite{WSBB}.  Moreover, as $\H_{\fl{N/2}}$ is exponentially large as
$N$ increases, it can be used as a resource for universal quantum computation.  For instance, we
can encode qubits as $\alpha |01\rangle + \beta |10\rangle$, thereby
performing universal quantum computation in $\H_{\fl{N/2}}$.  This is a
remarkable observation: we have found a system where quantum computation
can be achieved with a single switch, and where both the system and
control Hamiltonian are \emph{physical}, e.g. consist of nearest-neighbor two-body
interactions, which are very common in physics. It provides possibly the simplest and most elegant way of
achieving quantum computation so far (leaving efficiency issues beside
\cite{Burgarth}). Having only a single switch we avoid the experimental
difficulty of quickly changing field directions.

\subsection{XYZ chain}

For XYZ chain under $Z_1$ control with
\begin{subequations}
\label{eqn:xyz}
\begin{align}
  H_0&=\sum_j^N a_jX_jX_{j+1}+b_jY_jY_{j+1}+c_jZ_jZ_{j+1}\\
  H_1&=Z_1
\end{align}
\end{subequations}
where $a_j\ne b_j$, does the subspace controllability still exist? As
discussed in Example~1, there exists a parity symmetry $S_p=Z_1\cdots
Z_N$ satisfying $[H_0,S_p]=[H_1,S_p]=0$, with two invariant subspaces
$\H_1$ and $\H_{-1}$, corresponding to eigenvalues $\pm 1$ of $S_p$.  We
will show that the Hamiltonians cannot be further block-diagonalized on
each of the two subspaces, and the system is controllable on each of
them.  Notice that, compared to the XXZ chain, the number of invariant
subspaces for XYZ chain has reduced from $N+1$ to $2$, which is not too
surprising as we have broken the symmetry between X and Y directions
from XXZ to XYZ type, and some symmetries should disappear. In the
following we will identify all operators in $\L$ generated by $H_0$ and
$H_1$:
\begin{align*}
& [Z_1,H_0]\rightarrow a_1Y_1X_2-b_1X_1Y_2\rightarrow a_1X_1X_2+b_1Y_1Y_2\\
& [a_1Y_1X_2-b_1X_1Y_2,a_1X_1X_2+b_1Y_1Y_2]\\
&\rightarrow
(a_1^2+b_1^2)Z_1-2a_1b_1Z_2\rightarrow Z_2
\end{align*}
Continuing this process, we obtain all $Z_j$,
$a_jX_jX_{j+1}+b_jY_jY_{j+1}$, $a_jY_jX_{j+1}-b_jX_jY_{j+1}$ and
$H_2=\sum_j c_jZ_jZ_{j+1}$. Next, we have
\begin{align*}
  &[a_jY_jX_{j+1}-b_jX_jY_{j+1},Z_{j+1}]\rightarrow a_jY_jY_{j+1}-b_jX_jX_{j+1},
\end{align*}
and together with $a_jX_jX_{j+1}+b_jY_jY_{j+1}$ we can decouple and get
$X_jX_{j+1}$ and $Y_jY_{j+1}$. Similarly we can decouple and
independently generate $X_jY_{j+1}$ and $Y_jX_{j+1}$. This a major
difference from the XXZ case, where the XX and YY operators at
neighboring locations cannot be decoupled. Due to such decoupling, we
expect that the dynamical Lie algebra $\L$ generated by $H_0$ and $H_1$
will be larger than the XXZ case. Next, repeating the same generation
process by calculating the commutators, we get the following set series
$M_k$ of $k$-body operators:\\ (1) $M_1$: $Z_k$;\\ (2) $M_2$: $P_jP_k$,
where $P$ can be $X$ or $Y$, and $Z_jZ_k$;\\ (3) $M_3$: $P_jP_kZ_m$ and
$Z_kZ_mZ_n$\\ $\cdots\cdots\cdots$\\ ($\ell$) $M_\ell$: when $\ell$ is
even, we can generate:
\begin{align*}
&P_{m_1}P_{m_2}Z_{m_3}\cdots Z_{m_\ell}\\
&P_{m_1}P_{m_2}P_{m_3}P_{m_4}Z_{m_5}\cdots Z_{m_\ell}\\
& \cdots\cdots\cdots\\
&P_{m_1}P_{m_2}\cdots P_{m_\ell-1}P_{m_\ell}\\
&Z_{m_1}Z_{m_2}\cdots Z_{m_\ell}
\end{align*}
When $\ell$ is odd, we can generate:
\begin{align*}
&P_{m_1}P_{m_2}Z_{m_3}\cdots Z_{m_\ell}\\
&P_{m_1}P_{m_2}P_{m_3}P_{m_4}Z_{m_5}\cdots Z_{m_\ell}\\
& \cdots\cdots\cdots\\
&P_{m_1}P_{m_2}\cdots
P_{m_\ell-2}P_{m_\ell-1}Z_{m_\ell}\\
&Z_{m_1}Z_{m_2}\cdots Z_{m_\ell}
\end{align*}

Compared with XXZ chain, where we can only generate the coupled
Z-type operator, such as $(Z_1-Z_2)Z_3Z_4$, for XYZ chain, we can
separately generate $Z_1Z_3Z_4$ and $Z_2Z_3Z_4$. $M_\ell$ is be divided
into two subsets: the set of $P-Z$-type operators and the set of
$Z$-type operators, where each $P-Z$ operator can contain $2p$ number of
$P$'s and $N-2p$ number of $Z$'s, $p=1,\ldots,\fl{\ell/2}$. Hence,
following some basic combinatorics argument, we have:
\begin{align*}
  \rank(M_\ell)=\sum_{p=1}^k={2^{2p}}\binom{N}{2p}\fl{\ell/2}
  r_p\binom{N-2p}{\ell-2p} +\binom{N}{\ell},
\end{align*}
and the dimension of $\L$($N>2$):
\begin{align*}
\dim(\L)=\sum N_\ell =2^{N}\sum_{k=0}^{\fl {N/2}} \binom{N}{2k}-2=2^{2N-1}-2
\end{align*}
where we have used the identity
\begin{align*}
\sum_{k=0}^{\fl {N/2}} \binom{N}{2k}=\sum_{k=0}^{\fl {N/2}} \binom{N}{2k+1}=2^{N-1}
\end{align*}

Since $H_0$ and $H_1$ are simultaneously block-diagonalized on
$\H_1\oplus \H_{-1}$, $\L$ must be a subalgebra of
$\L_T=\uu(2^{N-1})\oplus \uu(2^{N-1})$.  Moreover, since the $k$-body
$Z$ operators in $\L$ are generated from the $(k+1)$-body $P-Z$
operators, $\L$ does not include two $Z$-type operators, the identity
$I$ and $S_p$, which are however included in $\L_T$. Hence, we have
\begin{align*}
\dim(\L)&\le \dim(\L_T)-2=2^{2N-1}-2=\dim(\L)
\end{align*}

Hence $\dim(\L)$ achieves the allowed maximal value, which is only true
when restricted on each of the subspace $\H_1$ and $\H_{-1}$,
$\L=\u(2^{N-1})$ or $\su(2^{N-1})$. Noticing that $H_0$ and $H_1$ are
trace-zero on $\H_1$ and $\H_{-1}$ for $N>2$, we must have
$\L=\su(2^{N-1})$ on both $\H_1$ and $\H_{-1}$ for $N>2$. When $N=2$, it
is easy to check that $\L=\u(2)$ on $\H_1$ and $\H_{-1}$. Thus, we have
proved the following theorem:

\begin{theorem}
\label{thm:full_control_k} For an XYZ chain of length $N$ with a single
local control on the end spin in $Z$ direction, the system is
controllable on each of the two invariant subspaces $\H_1$ and
$\H_{-1}$.
\end{theorem}

\subsection{When Control Has a Leakage on Neighboring Spins}

The previous assumption of control on a single spin only holds in
theory.  In practice, it is difficult to apply a control field that only
acts on a single spin without affecting its neighbors.  Hence, a more
realistic assumption is that the local end control has a leakage on the
neighboring spins with $H_1=\sum_{j=1}^{k} \gamma_jZ_j$. We consider
two common types of leakage: linear $\gamma_j=-\alpha j+\beta$ and exponential
$\gamma_j=e^{-\mu(j-1)^2}$ decays. In the following, we show that the
subspace controllability results discussed so far are robust against
such control leakage, i.e., when single control field $Z_1$ has some leakage
on the neighboring spins, the system is still controllable in the
invariant subspaces. Under the leakage assumption,
\begin{subequations}
\label{eqn:leakage}
\begin{align}
H_0&=\sum_{j=1}^N a_jX_jX_{j+1}+b_jY_jY_{j+1}+c_jZ_jZ_{j+1}\\
H_1&=\sum_{j=1}^{k} \gamma_jZ_j,
\end{align}
\end{subequations}
Defining adjoint action of $H_1$ on $H_1$ as $\Ad_{H_1}(H_0)=[H_1,H_0]$
and $A_j=a_jX_jX_{j+1}+b_jY_jY_{j+1}$, we have
\begin{align*}
\Ad_{H_1}^{(2)}(H_0)&=(\gamma_1-\gamma_2)^2 A_1+ \cdots + (\gamma_{k-1}-\gamma_k)^2 A_{k-1}+\gamma_k^2 A_k\\
\Ad_{H_1}^{(4)}(H_0)&=(\gamma_1-\gamma_2)^4 A_1+ \cdots + (\gamma_{k-1}-\gamma_k)^4 A_{k-1}+\gamma_k^4 A_k\\
\cdots\cdots &\cdots \cdots\cdots \\
\Ad_{H_1}^{(2k)}(H_0)&=(\gamma_1-\gamma_2)^{2k} A_1+ \cdots + (\gamma_{k-1}-\gamma_k)^{2k} A_{k-1}+\gamma_k^{2k} A_k
\end{align*}
where the coefficients in this expression can be denoted by
\begin{align*}
V=\begin{pmatrix}
(\gamma_1-\gamma_2)^2 &\cdots&(\gamma_{k-1}-\gamma_k)^2 & \gamma_k^2\\
(\gamma_1-\gamma_2)^4 &\cdots&(\gamma_{k-1}-\gamma_k)^4 & \gamma_k^4\\
\vdots & \cdots & \vdots & \vdots \\
(\gamma_1-\gamma_2)^{2k} &\cdots&(\gamma_{k-1}-\gamma_k)^{2k} &
\gamma_k^{2k}
\end{pmatrix}.
\end{align*}

(1) When the leakage of the local control is linear, i.e.,
$\gamma_j-\gamma_{j+1}=\gamma_{\ell}-\gamma_{\ell+1}$ for different $j$
and $\ell$, we can generate the operator
$A_k=a_kX_kX_{k+1}+b_kY_kY_{k+1}$ from any two rows of $V$. Analogously,
we can generate $a_kX_kY_{k+1}-b_kY_kX_{k+1}$ and hence generate
$Z_k-Z_{k+1}$. From $A_k$, we can also generate
$B_0=\Ad_{H_1}^{(2)}(H_0)-\gamma_k^2 A_k$. From $Z_k-Z_{k+1}$ and $B_0$,
we can sequentially generate $A_j$ and $Z_j$, $j=k-1,k-2,\cdots,1$.

(2) When the leakage of the local control decays nonlinearly, e.g.,
$\gamma_j=e^{-\mu(j-1)^2}$, we have
$\gamma_j-\gamma_{j+1}\ne\gamma_{\ell}-\gamma_{\ell+1}\ne \gamma_k $,
and from the property of Vandermonde matrix, $\det(V)\ne 0$. Hence we
can generate each $A_j$, $j=1,\ldots,k$. Together with $H_1$, we can
decouple and generate $Z_j$, $j=1,\ldots,k$. 

Hence, in both cases, $\L$ generated by $H_0$ and $H_1$ in
(\ref{eqn:leakage}) is the same as that generated by $H_0$ and
$H_1=Z_1$. In general, for other types of nonlinear leakage, the above
reasoning is valid for \emph{almost} all choices of $\gamma_j$. Thus we
have:

\begin{theorem}
For an XXZ or XYZ chain of length $N$, under a single local control
on the end spin in Z-direction with leakage to the neighboring spins,
the system is controllable on each of the invariant subspaces.
\end{theorem}

\section{Minimal Controls for Full Controllability}
\label{sec:minimal}

In previous section, we have provided a complete discussion of the
control problem of spin chains with the least control degree of freedom,
i.e., a single local control at the end of the spin chain.  In general,
as the number of controls increases, existing symmetries will disappear
and the system will become fully controllable on the entire Hilbert
space under a sufficient number of independent controls.  Therefore,
another interesting theoretical question is to ask when such transition
happens from an uncontrollable system to a fully controllable
one.  Alternatively, we can ask what are the minimal controls that can
make the chain fully controllable, which is the main topic of this
section. We will base on the results in previous discussions and add
more controls to the control systems under (\ref{eqn:xxz}) or
(\ref{eqn:xyz}).

\subsection{Controlling $Z_1$ and $X_1$}

In~\cite{Burgarth-Bose}, it was proved by the propagation property that
an XXZ chain with two independent controls $H_1=Z_1$ and $H_2=X_1$ is
fully controllable on the entire space.  We can rederive this result
from our analysis in previous section: observing the operators generated
by $H_0$ and $H_1$, and writing down the operators generated by $H_0$
and $H_2$, it is easy to see that we can generate all $k$-body Pauli
operators, $k=1,\ldots,N$ in $\uu(2^N)$. Hence the system is fully
controllable.

\begin{theorem}
\label{thm:end_spin_full} For an XXZ or XYZ chain of length $N$,
with two local controls on the end spin, $H_1=Z_1$ and $H_2=X_1$, the
system is controllable on the whole space.
\end{theorem}

\subsection{Controlling $Z_k$ and $X_k$}

In Theorem~\ref{thm:end_spin_full}, we have shown that if we can fully
control the end spin, then the system is controllable on the whole
space.  What if we can fully control one spin at other locations?  We
will prove that for a general XYZ chain two independent controls
on the $k$th spin in Z and X directions
\begin{align*}
  H_0&= \sum_j^N a_jX_jX_{j+1}+b_jY_jY_{j+1}+c_jZ_jZ_{j+1}\\
  H_1&= Z_k,\quad H_{2}=X_k
\end{align*}
are sufficient for controllability on the whole space, except when
$N=2k+1$, where the Hamiltonians exhibit a mirror permutation symmetry
with respect to the center $k$th spin.  Specifically, for $k\le
\fl{N/2}$, let us calculate the operators in $\L$ generated by the three
Hamiltonians.

\begin{align*}
&[Z_k,H_0]\\
 \rightarrow &(a_{k-1}X_{k-1}Y_k-b_{k-1}Y_{k-1}X_k)+
(a_kY_kX_{k+1}-b_kX_kY_{k+1})\\
\rightarrow &(a_{k-1}X_{k-1}X_k+b_{k-1}Y_{k-1}Y_k)+
(a_kX_kX_{k+1}+b_kY_kY_{k+1})\\
\rightarrow
&(a_{k-1}^2+b_{k-1}^2)Z_k-2a_{k-1}b_{k-1}Z_{k-1}\\
&+(a_k^2+b_k^2)Z_k-2a_kb_kZ_{k+1}\\
&+2(a_{k-1}a_kX_{k-1}X_{k+1}+b_{k-1}b_kY_{k-1}Y_{k+1})Z_k\equiv P_k\\
&[X_k,[X_k,P_k]]-P_k\rightarrow
d_{k-1}Z_{k-1}+d_{k+1}Z_{k+1}\equiv Q_1\\
&[Q_1,H_3]\rightarrow\cdots\rightarrow d_{k-2}Z_{k-2}+d_{k+2}Z_{k+2}\equiv Q_2
\end{align*}

Continuing this process, we can sequentially generate:
$d_{k-1}Z_{k-1}+d_{k+1}Z_{k+1}$, $d_{k-2}Z_{k-2}+d_{k+2}Z_{k+2}$,
$\cdots$, $d_1Z_1+d_{2k-1}Z_{2k-1}$,
$Z_{2k}-Z_{2k-1}$,
$Z_{2k+1}-Z_{2k}$,
$\cdots$,$Z_N-Z_{N-1}$.
%\begin{align*}
%&d_{k-1}Z_{k-1}+d_{k+1}Z_{k+1}\\
%&d_{k-2}Z_{k-2}+d_{k+2}Z_{k+2}\\
%&\cdots\cdots\cdots\\
%&d_1Z_1+d_{2k-1}Z_{2k-1}\\
%&Z_{2k}-Z_{2k-1}\\
%&Z_{2k+1}-Z_{2k}\\
%&\cdots\cdots\cdots\\
%&Z_N-Z_{N-1}
%\end{align*}
as well as
\begin{align*}
&(a_{k-1}X_{k-1}X_k+b_{k-1}Y_{k-1}Y_k)+
(a_kX_kX_{k+1}+b_kY_kY_{k+1})\\
&\cdots\cdots\cdots\\
&(a_1X_1X_2+b_1Y_1Y_2)+
(a_{2k-1}X_{2k-1}X_{2k}+b_{2k-1}Y_{2k-1}Y_{2k})\\
&a_{2k-1}X_{2k-1}X_{2k}+b_{2k-1}Y_{2k-1}Y_{2k}\\
&a_{2k}X_{2k}X_{2k+1}+b_{2k}Y_{2k}Y_{2k+1}\\
&\cdots\cdots\cdots\\
&a_{N-1}X_{N-1}X_N+b_{N-1}Y_{N-1}Y_N
\end{align*}
Then starting from $[Z_{N-1}-Z_N, H_0]$, we can sequentially generate
$Z_{N-1}-Z_N$, $Z_{N-2}-Z_{N-1}$, $\cdots$, $Z_{k}-Z_{k+1}$.  Together
with $Z_k$, we can generate $Z_{k+1}$, and hence $Z_{k+1},\ldots,
Z_{2k-1},\ldots,Z_N$. Then together with $d_1Z_1+d_{2k-1}Z_{2k-1}$, we
can decouple and generate $Z_1$.  Analogously, starting from
$[X_k,H_0]$, we can generate $X_1$. Hence by
Theorem~\ref{thm:end_spin_full}, we have:

\begin{theorem}
\label{thm:k_spin_full} For an XXZ or XYZ chain of length $N$, with
two local controls on the $k$-th spin, $H_1=Z_k$ and $H_2=X_k$, and
$N\neq 2k+1$ the system is controllable on the whole space.
\end{theorem}

\subsection{Controlling $Z_1$ and $X_k$}

Next we will see that two independent controls $Z_1$ and $X_k$ on the
first and the $k$th spins are also enough for full controllability for a
XYZ or XXZ chain, with Hamiltonians,
\begin{align*}
 H_0 &= \sum_j^N a_jX_jX_{j+1}+b_jY_jY_{j+1}+c_jZ_jZ_{j+1}\\
 H_1 &= Z_1,\quad H_{2}=X_k.
\end{align*}
Earlier we showed that $H_0$ and $H_1$ can generate 
\begin{align*}
[Z_1,H_0] \rightarrow &Z_j\\
\rightarrow &a_jX_jX_{j+1}+b_jY_jY_{j+1}\\
\rightarrow &a_jX_jY_{j+1}-b_jY_jX_{j+1}\\
\rightarrow & \sum_j^N c_jZ_jZ_{j+1}\equiv H_3
\end{align*}
and subsequently
\begin{align*}
&[X_k,H_3] \rightarrow c_{k-1}Z_{k-1}Y_k+ c_kY_kZ_{k+1}\\
&[X_k,a_{k-1}X_{k-1}X_k+b_{k-1}Y_{k-1}Y_k]\rightarrow Y_{k-1}Z_k\\
&[X_k,Y_{k-1}Z_k]\rightarrow Y_{k-1}Y_k\\
&[Y_{k-1}Y_k,c_{k-1}Z_{k-1}Y_k+ c_kY_kZ_{k+1}]\rightarrow X_{k-1}
\end{align*}
Continuing such process, we can sequentially generate $X_{k-2}$,
$X_{k-3}$, $\cdots$, $X_2$, $X_1$. By Theorem~\ref{thm:end_spin_full},
we have
\begin{theorem}
\label{thm:xyz_1kj} For an XXZ or XYZ chain of length $N$, with two
local controls $Z_1$ and $X_k$, $1\le k\le N$, on the first and the
$k$-th spins, the system is controllable on the whole space.
\end{theorem}

\section{Internal Symmetries and XX Chains}
\label{sec:xx}

In previous sections we have discussed the controllability results for
XXZ or XYZ chains with one or two local controls.  The fact that there
are spin-spin interactions in all three directions is essential to
guarantee subspace controllability or full controllability, and we
expect that this will no longer be true if the spin chain is reduced to
XX, XY or Ising types, as absence of the ZZ terms in $H_0$ allows for more
symmetries.  Since the XX chain is particularly important as a
theoretical model, we first investigate how the controllability properties 
change in this case.

We start again with an XX chain and a single end control in the
Z-direction
\begin{align}
\label{eqn:xx}
H_0=\sum_{n}^N \gamma_n(X_nX_{n+1}+Y_nY_{n+1}), \, H_1=Z_1
\end{align}

\begin{theorem}
\label{thm:XX1} For an XX chain of length $N$ with a single local
control on the end spin in Z-direction, the associated dynamical Lie
algebra $\L$ is a faithful representation of $\u(N)$ in each excitation
subspace $\H_k$, $k=1,\ldots,N-1$.
\end{theorem}

\begin{IEEEproof}
In order to prove $\L$ is indeed isomorphic to $\u(N)$, we determine all
independent operators generated from the Hamiltonians, as we did in
previous sections, and show that these operators satisfy the same
commutation relations as the standard basis of
$\u(N)$~\cite{WS_symmetry}.  However, to make the following analysis of
faithful representations simpler and more convenient, we transform the
original representation using the Jordan-Wigner(JW) transformation, a
powerful tool initially developed in theoretical physics: defining
\begin{align*}
  a_{m}:=\prod_{n<m}Z_{n}\left(X- iY\right)_{m}
\end{align*}
where $a_m$ are fermionic annihilation operators, with the canonical
anticommutation relations
\begin{align*}
  \{a_m, a_n\}      &= \{a_m^\dag, a_n^\dag \} = 0,  &\quad
  \{a_m, a_n^\dag\} &= \delta_{mn},
\end{align*}
the Hamiltonians are transformed into:
\begin{align*}
  H_0 &= \sum_{n=1}^N \gamma_{n} (a_n^\dag a_{n+1} + a_{n+1}^\dag a_n)\\
  H_1 &= a_1^\dag a_1
\end{align*}
By calculating the commutation relations between the Hamiltonians, we
can verify the following identities:
\begin{align*}
 [iH_1,iH_0]     & \to y_{12} := a_1^\dag a_2 - a_2^\dag a_1 \\
 [iH_1,y_{12}]   & \to x_{12} := i(a_1^\dag a_2 + a_2^\dag a_1)\\
 [x_{12},y_{12}] & \to z_2 := i a_2^\dag a_2
\end{align*}
Then we can generate $\bar H_0= H_0-\gamma_1 x_{12}$, which represents
the system Hamiltonian for a chain of length $N-1$, and $z_2$ amounts to
the end control on $\bar H_0$.  Thus by induction we can sequentially
generate
\begin{align*}
  y_{n,n+1} &:= a_n^\dag a_{n+1} - a_{n+1}^\dag a_n, & n=1,\ldots, N-1 \\
  x_{n,n+1} &:= i(a_n^\dag a_{n+1} + a_{n+1}^\dag a_n), & n=1,\ldots, N-1 \\
  z_{n}     &:= i(a_n^\dag a_n),                        & n=1,\ldots N.
\end{align*}
Hence, we have generated three kinds of anti-Hermitian operators:
$x_{n,n+1}$, $y_{n,n+1}$ and $z_n$, satisfying
\begin{subequations}
\begin{align}
 [x_{mn},z_n]      &= y_{mn}\\
 [y_{mn},z_n]      &=-x_{mn}\\
 [x_{mk},x_{kn}]&= y_{mn}\\
 [x_{mk},y_{kn}]&=-x_{mn}\\
 [x_{mn},y_{mn}]   &= 2(z_m-z_n)
\end{align}
\end{subequations}
which are the same as the commutation relations satisfied by the
standard basis of $\u(N)$~\cite{WS_symmetry}.  Hence, we have
$\L=\uu(N)$.

Finally, in order to show $\L$ is a faithful representation in each
excitation subspace $\H_k$, $k=1,\cdots, N-1$, it is sufficient to show
that the images of the generators $x_{n,n+1}$, $y_{n,n+1}$ and $z_n$ in
$H_k$ are nonzero. Choose a vector $\ket{\alpha}\in \H_k$ in the
computational basis with $n$ and $n+1$ positions as $\ket{0}$ and
$\ket{1}$, and define another basis vector $\ket{\beta}\in \H_k$ such
that $\beta$ only differs from $\alpha$ at $n$ and $n+1$ positions, with
values $\ket{1}$ and $\ket{0}$. Restricted on $\H_k$, we have
\begin{align*}
\bra{\beta}y_{n,n+1}\ket{\alpha}&=\bra{\beta} a_n^\dag a_{n+1} \ket{\alpha}= 1 \\
\bra{\beta}x_{n,n+1}\ket{\alpha}&=\bra{\beta} a_n^\dag a_{n+1} \ket{\alpha}= i \\
\bra{\beta}z_{n}\ket{\beta}&=\bra{\beta} a_n^\dag a_{n} \ket{\beta}= 1 
\end{align*}
Hence, on each $\H_k$, $k=1,\cdots, N-1$, $\L$ is a faithful
representation of $\u(N)$.
\end{IEEEproof}

The above theorem implies that for XX chains with a $Z_1$ end control
the system is controllable only on the single excitation subspace (or
its mirror image, the $N-1$ excitation subspace).  In order to derive
full controllability we require more controls.  If we add another local
control $X_1$, we have the following result (proof in~\cite{Thomas}):

\begin{theorem}
\label{thm:XX3} For an XX chain of length $N$ with two local controls
on the end spin, $H_1=Z_1$ and $H_2=X_1$, the system is not controllable
on the full Hilbert space and $\L=\so(2N+1)$.
\end{theorem}

To better understand this result we find all operators in $\L$ and the
identify their symmetries.  According to previous discussions, the
system Hamiltonian of the XX chain, $H_0$, and $Z_1$ generate
$X_kX_{k+1}+Y_kY_{k+1}$, $X_kY_{k+1}-Y_kX_{k+1}$ and $Z_k$.  On the
other hand, $H_0$ and $X_1$ generate $Z_1Y_2$ and then $Y_1Y_2$ and
$X_1X_2$.  Similarly we can decouple and generate $X_1Y_2$ and $Y_1X_2$.
Also $X_1$ and $Z_1$ will generate $Y_1$, and together with $H_0$ we can
generate $Z_1X_2$.  Continuing this process, we have
\begin{align*}
  &[Z_1X_2,X_2X_3+Y_2Y_3] \to Z_1Z_2Y_3 \\
  &[Z_1Z_2Y_3,Z_3] \to Z_1Z_2X_3, \,\cdots\cdots\cdots \\
  &[Z_1\cdots Z_{N-2}X_{N-1},X_{N-1}X_N+Y_{N-1}Y_N] \to Z_1\cdots Z_{N-1}Y_N \\
  &[Z_1\cdots Z_{N-1}Y_N,Z_N] \to Z_1\cdots Z_{N-1}X_N\\
  &[Z_1\cdots Z_{j-1}X_j,Z_1\cdots Z_{k-1}X_k] \to X_jZ_{j+1}\cdots Z_{k-1}X_k\\
  &\to X_jZ_{j+1}\cdots Z_{k-1}Y_k \to Y_jZ_{j+1}\cdots Z_{k-1}X_k
\end{align*}
Hence we can generate $4\binom{N}{2}$ operators of the form
$P_jZ_{j+1}\cdots Z_{k-1}P_k$, where $P$ can be either $X$ or $Y$, and
there are $N$ operators $Z_1\cdots Z_{k-1}P_k$, where $P=X$ or $Y$.
Together with $N$ number of $Z_k$, we have $\dim(\L)=4\binom{N}{2}+2N+N
=N(2N+1)=\dim\so(2N+1)$.  This suggests that the Lie algebra is a
representation of $\so(2N+1)$, although since the Lie algebra is not the
maximal Lie algebra on the subspace, calculating the dimension is not
sufficient.  However, setting $S=Y_1X_2Y_3X_4\cdots$, it is easy to
verify that we have $H_j^TS+SH_j=0$ for $j=0,1,2$ and thus $S$ is an
internal symmetry.  Depending on the number of $Y$'s in $S$, $S$ is
either symmetric or antisymmetric and thus defines an orthogonal or
sympletic symmetry.  This means that $\L$ is a subalgebra of $\so(2^N)$
or $\sp(2^{N-1})$, respectively, and therefore not controllable.
Strictly, this is still not sufficient to conclude that $\L\simeq
\so(2N+1)$.  However, the table of irreducible simple subalgebras of
$\su(2^N)$ in~\cite{Thomas} suggests that $\so(2N+1)$ is indeed the only
possible subalgebra that has the correct dimension and symmetry.

Next, we study the XX chain subject to two controls, $Z_1$ and
$X_2$.  In this case, it is straightforward to verify that
\begin{align}
\label{eqn:internal_sym} S=X_1Y_2X_3Y_4\cdots
\end{align}
defines an internal symmetry. Depending on the number of $Y$'s in $S$,
denoted by $N_Y$, the symmetry can be either orthogonal or symepectic:
if $N_Y$ is even, $S$ is an orthogonal symmetry, and if $N_Y$ is odd,
$S$ is a symplectic symmetry.

\begin{theorem}
\label{thm:xx_z1x2} For an XX chain of length $N$ subject to two local
controls $H_1=Z_1$, $H_2=X_2$, the dynamical Lie algebra 
\begin{equation}
  \L = \begin{cases}
	\so(2^N)     & N \in \{4k, 4k+1: k\ge 1\}, \\
                            \sp(2^{N-1}) & N \in \{4k+2, 4k+3: k\ge 0\}.
	\end{cases} 
\end{equation} 
\end{theorem} 

By the definition of internal symmetry, any operator $M\in \L$ also
satisfies the symmetry: $M^TS+SM=0$.  Define the symmetry-preserving set
$\G$ to be the set of all general Pauli operators $M=P_1P_2\cdots P_N$,
where $P=X,Y,Z$, satisfying $M^TS+SM=0$.  We have $\L\subset \G$ and
$\Span(\G)\subset \su(2^N)$ forms a Lie algebra.  In the following, we
determine $\rank(\G)$ and show that all operators in $\G$ can be
generated from the Hamiltonians, i.e., $\L=\Span(\G)$.  Denote by $N_e$
($N_o$) the number of even (odd) positions of the XX chain; by $n_e$ 
($n_o$) the number of $X$'s and $Y$'s contained in $M$ at the even (odd) 
positions, and $n_Z$ as the number of $Z$'s contained in $M$.  We have
$N=N_e+N_o$, $0 \le n_e \le N_e$, and $0 \le n_o \le N_o$.

Here is the rule for any $M \in \G$: if $n_e$ is odd, then $n_Z$ is
even; if $n_e$ is even, then $n_Z$ is odd. For example, $X_1Z_2$, and
$X_1X_2X_3$ both preserve the symmetry, while $Z_1X_2$ and $X_1Y_2Z_3$
do not. Notice that the $X$'s and $Y$'s contained in $\M$ are
interchangeable at the same location to preserve the symmetry. For
example, $Y_1X_2X_3$ and $Y_1Y_2X_3$ are also in $\G$.  This fact is
useful when evaluating $\rank(\G)$: for a given $M \in \G$, containing
$p$ number of $X$'s and $Y$'s, we can write down $2^p$ number
independent operators in $\G$ based on $M$.

Specifically, for $N=2k$, $S=X_1Y_2X_3Y_4\cdots X_{N-1}Y_N$ and
$N_e=N_o=k$. If $n_e=0$, then $n_Z$ can be $1,3,\ldots, k$, and $n_o$
can be chosen as $\binom{N_o}{0}$, $\binom{N_o}{1}$, $\binom{N_o}{2}$,
$\ldots$. Denoting $N_{n_e=m}$ as the number of $M \in \G$ with $n_e=m$,
$N_{n_e=0}$ is equal to:
\begin{align*} 
&\textstyle
  2^0 \binom{N_e}{0}\Big[ 2^0\binom{N_o}{0}\left[ \binom{N}{1}+\binom{N}{3}+\cdots\right]\\
& \textstyle
  \qquad+2^1\binom{N_o}{1}\left[\binom{N-1}{1}+ \binom{N-1}{3}+\cdots\right]\\
&\textstyle
  \qquad+2^2\binom{N_o}{2}\left[\binom{N-2}{1}+ \binom{N-2}{3}+\cdots\right]\\
&\textstyle\qquad
  +\cdots +2^k\binom{N_o}{k}\left[\binom{N-k}{1}+ \binom{N-k}{3}+\cdots \right] \Big]\\
=&\textstyle
  \binom{N_e}{0} 2^{N-1} \left[ \binom{N_o}{0}+\binom{N_o}{1}+\cdots \right]=\binom{N_e}{0} 2^{N-1} 2^{N_o}
\end{align*}
Analogously, for $n_e=1$, $n_Z$ can be $0,2,4,\ldots$ and $N_{n_e=1}$
is:
\begin{equation*}
\textstyle
   \binom{N_e}{1} 2^{N-1} \left[ \binom{N_o}{0}+\binom{N_o}{1}+\cdots+\binom{N_o}{k}\right]
 = \binom{N_e}{1} 2^{N-1} 2^{N_o}
\end{equation*}
Repeating such process, we find $N_{n_e=m}=\binom{N_e}{m} 2^{N-1}
2^{N_o}$.  Finally, for $n_e=k$, if $k$ is even, then $n_Z$ should be
odd, and $N_{n_e=k}$ is equal to:
\begin{align*}
 &\textstyle
 2^k \binom{N_e}{k}\Big[ 2^0\binom{N_o}{0}\left[ \binom{N-k}{1}+\cdots+ \binom{N-k}{N-k-1}\right]\\
 &\textstyle
 \qquad+2^1\binom{N_o}{1}\big[\binom{N-k-1}{1}+ \cdots+ \binom{N-k-1}{N-k-1}\big]\\
 &\textstyle\qquad
 +\cdots +2^{k-1}\binom{N_o}{k-1}\binom{1}{1} \Big] \\
=&\textstyle
  \binom{N_e}{k} 2^{N-1} \left[ \binom{N_o}{0}+\binom{N_o}{1}+\cdots+\binom{N_o}{k-1}  \right]\\
=&\textstyle
  \binom{N_e}{k} 2^{N-1}( 2^{N_o}-1)
\end{align*}
Therefore, 
\begin{align*}
\rank(\G)
&= \textstyle \sum_{m=0}^{k}N_{n_e=m}\\
&= \textstyle 2^{N-1} 2^{N_o} \left[\binom{N_e}{0}+\binom{N_e}{1}+\cdots+\binom{N_e}{k}\right]-2^{N-1}\\
&= \textstyle 2^{2N-1}-2^{N-1}  =\dim\left(\so(2^N)\right)
\end{align*}

If $k$ is odd, then $n_Z$ should be even, and $N_{n_e=k}$ is:
\begin{align*}
 &\textstyle
  2^k \binom{N_e}{k}\Big[ 2^0\binom{N_o}{0}\big[ \binom{N-k}{0}+\cdots+ \binom{N-k}{N-k-1}\big]\\
 &\textstyle\qquad
  +2^1\binom{N_o}{1}\big[\binom{N-k-1}{0}+ \cdots+ \binom{N-k-1}{N-k-1}\big]\\
 &\textstyle\qquad
  +\cdots +2^{k-1}\binom{N_o}{k-1}\binom{1}{0}+2^{k}\binom{N_o}{k}\binom{0}{0}\Big] \\
=&\textstyle
  \binom{N_e}{k} 2^{N-1} \Big[ \binom{N_o}{0}+\binom{N_o}{1}+\cdots+\binom{N_o}{k-1}+2  \Big]\\
=&\textstyle
  \binom{N_e}{k} 2^{N-1}( 2^{N_o}+1)
\end{align*}
Therefore,  
\begin{align*}
\rank(\G)=\sum_{m=0}^{k}N_{n_e=m}= 2^{2N-1}+2^{N-1}=\dim\big(\sp(2^{N-1})\big)
\end{align*}

For $N=2k-1$, we have $N_o=k$ and $N_e=k-1$. Analogously, for both even
and odd $k$ cases, we can respectively find the value of
$\rank(\G)$. Specifically, for odd $k$, and $\rank(\G)=
2^{2N-1}+2^{N-1}=\dim\big(\sp(2^{N-1})\big)$; for even $k$, $\rank(\G)=
2^{2N-1}-2^{N-1}=\dim\big(\so(2^N)\big)$. Hence we have the following
result:
\begin{lemma}
For an XX chain of length $N$, with two local controls $H_1=Z_1$,
$H_2=X_2$, when $N=4k$ or $4k+1$, $k\ge 1$,
$\rank(\G)=\dim\big(\so(2^N)\big)$; when $N=4k+2$ or $4k+3$, $k \ge 0$,
$\rank(\G)=\dim\big(\sp(2^{N-1})\big)$.
\end{lemma} 

Next, in order to prove Theorem~\ref{thm:xx_z1x2}, it is sufficient to
show that $\L=\Span(\G)$. Then, since $\L$ is a subalgebra of $\so(2^N)$
or $\sp(2^{N-1})$ and $\dim(\L)=\rank(\G)=\dim\big(\so(2^N)\big)$ or
$\dim\big(\sp(2^{N-1})\big)$, $\L$ must be $\so(2^N)$ or $\sp(2^{N-1})$.
\begin{align*}
&[H_0,Z_1] \to Z_k, k=1,\ldots,, N\\
&\to X_kX_{k+1}+Y_kY_{k+1},  X_kY_{k+1}-Y_kX_{k+1}\\
&[X_1X_2+Y_1Y_2,X_2]\to Y_1Z_2 \to Y_1Y_2 \to X_1X_2\\ 
&[X_1Y_2-Y_1X_2,X_2]\to X_1Z_2 \to X_1Y_2 \to Y_1X_2
\end{align*}
Thus, with the help of $X_2$, we can decouple and generate $P_1Q_2$ and
similarly $Q_2P_3$, where $P=X,Y$ and $Q=X,Y,Z$.  Then,
$[X_2X_3,X_3X_4+Y_3Y_4]\to X_2Z_3Y_4$, $[X_2X_3, X_2Z_3Y_4] \to
Y_3Y_4\to X_3X_4 \to\cdots \to Y_{N-1}Y_{N} \to X_{N-1}X_N$.  And
similarly we can decouple and generate all $X_kY_{k+1}$ and
$Y_kX_{k+1}$. Hence we can generate all $P_kP_{k_1}$, where $P=X,Y$. We
can see that it is very important to first generate a single decoupled
$X_1X_2$, and then using it we can decouple all the other
$P_kP_{k+1}$. This is possible only when we have $X_2$ initially. Hence,
this availability of $X_2$ control is \emph{essential} for
Theorem~\ref{thm:xx_z1x2}. Notice that as we can generate all $Z_k$, for
any $M\in \L$ that contains $X$ at the $k$th position, i.e., $M=\cdots
X_k\cdots$, we can also generate $[M,Z_k]\to M_s=\cdots
Y_k\cdots$. Hence, it is sufficient to show the result for $M$ only
contain $X$ and $Z$ operators.  From $X_1Z_2$:
\begin{align*}
&[X_1Z_2, Y_2X_3]\to X_1X_2X_3, \, &[ X_1X_2X_3, X_2Y_3] \to X_1Z_3\\
&[X_1Z_3, Y_3X_4]\to X_1X_3X_4, \, &[ X_1X_3X_4, X_3Y_4] \to X_1Z_4\\
&\cdots \cdots \cdots \to X_1Z_{N}\\
&[ X_1X_2X_3, Y_1X_2] \to Z_1X_3,\, &[Z_1X_3, Y_1X_3X_4]\to X_1X_4
\end{align*}
Therefore, we can generate all $X_1Z_k$, $k=2,\ldots, N$, and $X_1X_4$. Next,
\begin{align*}
&[X_1X_3X_4,Z_1X_4]\to Y_1X_4 \to X_1X_4\\
&[X_1Z_4,X_1X_4] \to Y_4 \to X_4
\end{align*}
This means that from $X_2$, we can generate $X_4$. Then we can imagine
that the chain starts from the position $3$, and we control $Z_3$ and
$X_4$, and analogously we can generate $X_6$. This observation is very
important, as it tell us whenever we can generate an operator with
respect to positions $1$ and $2$, we can automatically write down the
analogous operators we can generate at positions $3$ and $4$, or $5$ and
$6$, etc.  For example, for an XX chain with length $N=20$, since we
have shown that we can generate $X_1X_4$, we immediately know that we
can also generate $X_3X_6$, $X_5X_8$, etc.

Now we are ready to prove Theorem~\ref{thm:xx_z1x2} by induction. For
$N=2$, it is trivial to show $\L=\Span(\G)=\sp(2)$; for $N=3$, we first
list all elements of $\G$ in Table~\ref{table1}.
\begin{table}
\begin{tabular}{|c|c|}
\hline
 $1$-body   & $Z_k$, $X_2$  \\ \hline
 $2$-body   & $X_1Z_k$, $X_3Z_k$, $X_kX_{k+1}$ \\\hline
 $3$-body   & $Z_1Z_2Z_3$, $Z_1X_2Z_3$, $X_1Z_2X_3$, $X_1X_2X_3$ \\\hline
\end{tabular}
\caption{All symmetry preserving operators in $\G$ for $N=3$.} \label{table1}
\end{table}

In Table~\ref{table1}, besides the operators we have already generated
as above, we can also get:
\begin{align*}
&[Z_1X_3,Z_2Y_3]\to Z_1Z_2Z_3\, &[Z_1Z_2Z_3, Y_2] \to Z_1X_2Z_3\\
&[X_1X_2, Y_2X_3] \to X_1Z_2X_3\, &[X_1Z_2X_3, Y_2] \to X_1X_2X_3
\end{align*}
Thus, we have generated all symmetry-preserving operators for $N=3$, and
hence $\L=\Span(\G)=\sp(2^2)$. Similarly, for $N=4$, we can generate all
symmetry-preserving operators in $\G$. In particular, for $2$-body
operators, we also need to show we can generate $X_3Z_4$, which is true,
as
\begin{align*}
&[Z_1X_3,Y_1Z_4]\to X_1X_3Z_4\\
&[X_1X_3Z_4, X_1Z_3] \to Y_3Z_4\to X_3Z_4
\end{align*}
For all other $3$-body and $4$-body operators, we can sequentially
generate them from existing $2$-body and $3$-body operators. For
example, we have
\begin{align*}
&X_1X_2X_3,Y_3X_4]\to X_1X_2Z_3X_4\to  X_1Z_2Z_3X_4 \\
&[X_1Z_2Z_3X_4, Y_1Z_4] \to Z_2Z_3X_4
\end{align*}
We can verify that all operators in $\G$ can be generated for $N=4$, and
$\L=\G=\so(2^4)$.

Next, assuming that for all $N\le 2k$, we can generate all operators in
$\G$, we aim to show that for $N=2k+1$ and $N=2k+2$, we can also
generate $\G$. First, when $N=2k+1$, for any $M=P_1P_2\cdots
P_{2k}P_{2k+1} \in \G$, if $P_{2k+1}$ is the identity, then $M$ is also
a symmetry-preserving operator for the chain with length $2k$, and by
induction assumption, $M$ can be generated. If $P_{2k+1}$ is not the
identity, but $P_1$ and $P_2$ are, then as discussed earlier, $M$ can be
considered as a symmetry-preserving operator for the XX chain starting
from positions $3$ to $N$, with length $2k-1$, and by induction
assumption, it can also be generated. The remaining case is when $M$ is
an operator where both $P_{2k+1}$ and $P_1$ or $P_2$ are not identity,
then we can always choose a symmetry-preserving operator
$M_s=P_1P_2Q_3Q_4$ from positions $1$ to $4$ such that $Q_3Q_4$ is
different from $P_3P_4$. For example, if $M=X_1Z_2X_3X_4\cdots$, then we
can choose $M_s=X_1Z_2X_3$. Defining $L=[M,M_s]\ne 0$, $L$ is a also
symmetry-preserving operator for the XX chain from positions $3$ to $N$,
and $M$ can be generated from $L$ and $M_s$, which can both be generated
by the induction assumption. Thus, we can generate all operators in $\G$
for $N=2k+1$. For $N=2k+2$, we repeat the same induction argument again
and derive the same result. Thus, we have completed the whole proof of
Theorem~\ref{thm:xx_z1x2}.

Theorem~\ref{thm:XX3} and Theorem~\ref{thm:xx_z1x2} strongly suggest
that for an XX chain, two independent local controls are not enough
for full controllability. Moreover, we can numerically check for a given
chain length $N$ that two local controls with any configurations do not
induce full controllability, i.e., for two-control problem, we will
never find a full controllability result that is true for arbitrary
$N$. Nevertheless, it is worthwhile to notice that when $N=4k+2$ or
$4k+3$ and $\L=\sp(2^{N-1})$, the system is pure-state
controllable~\cite{pure_control}, i.e., we can generate arbitrary target
pure state from the initial pure state under these controls.  Pure-state
controllability can be useful for many practical applications.

Finally, we find that three local controls are enough to make the XX
chain fully controllable:
\begin{theorem}
An XY or XX chain of length $N$ is fully controllable under three
local controls $Z_1$, $X_1$ and $X_2$ on the first and the second spins.
\end{theorem}

\begin{IEEEproof}
From previous discussion, $Z_1$, $X_1$ and $H_0$ for XY or XX chain
will generate $Z_j$, $j=1,\ldots, N$, $P_1P_2$, $P=X,Y,Z$, and
$a_jX_jX_{j+1}+b_jY_jY_{j+1}$. Together with $X_1$ and $X_2$ we can
further generate $Z_1X_j$, $Z_1Y_j$, and $Z_1Z_2$. We have
$[Z_1X_2,a_2X_2Y_3-b_2Y_2X_3] \to Z_1 Z_2X_3$,
$[X_1X_2,a_2X_2X_3+b_2Y_2Y_3] \to X_1 Z_2Y_3$, $[Y_1Z_2,X_1Z_2Y_3] \to
Z_1Y_3$, $ [Z_1Z_2X_3,Z_1Y_3] \to Z_2 Z_3$, $[Y_2Y_3,X_2] \to Z_2Y_3$,
$[Z_2Y_3,Z_2Z_3] \to X_3$.
%\begin{align*}
%  &[Z_1X_2,a_2X_2Y_3-b_2Y_2X_3] \to Z_1 Z_2X_3\\
% &[X_1X_2,a_2X_2X_3+b_2Y_2Y_3] \to X_1 Z_2Y_3\\
% &[Y_1Z_2,X_1Z_2Y_3] \to Z_1Y_3\\
% & [Z_1Z_2X_3,Z_1Y_3] \to Z_2 Z_3 \\
%&[Y_2Y_3,X_2] \to Z_2Y_3\\
%&[Z_2Y_3,Z_2Z_3] \to X_3
%\end{align*}
Hence, from $H_0$, $Z_1$, $X_1$ and $X_2$, we can first generate
$P_1P_2$, and then $X_3$ and $Z_2Z_3$. Then together with $Z_3$, we can
generate all $P_2P_3$, $P=X,Y,Z$.  Continuing such process, we can
sequentially generate $P_jP_{j+1}$. Then the following procedure is
exactly the same as the XYZ chain case with end controls $Z_1$ and
$X_1$, and will generate all $k$-body Pauli operators and the system is
controllable, with $\L=\su(2^N)$.
\end{IEEEproof}

\section{Conclusion}

In this paper we have studied two kinds of Lie algebra symmetries,
external and internal, and applied them to analyze controllability
problems for spin chains.  Specifically, for XXZ and XYZ chains under
single end local control, the Hilbert space decomposes into a set of
invariant subspaces and we have shown that the system is controllable on
each of these invariant subspaces. These models are arguably the
simplest type of quantum computers one can think of: a physical
Hamiltonian with a single switch giving rise to an exponentially large
dynamical Lie algebra.  We have also shown that this result stills holds
when there is control leakage effecting neighboring spins, implying the
robustness of this controllability result against control leakage which
is common for practical control systems.  

We have also addressed the question of minimal control resources for
full controllability on the entire space.  For various cases, we find
that two independent local controls are sufficient to make the system
fully controllable for XXZ and XYZ chains.  Finally, we have studied the
effect of eliminating the ZZ-coupling that reduces the XXZ chain to an
XX chain.  We find that in this case for a single local end control, the
dynamical Lie algebra is the same on each invariant subspace, and hence
the system is only controllable in the first excitation subspace.
Moreover, through the investigation of the internal symmetries, we have
discussed the two-control problem of an XX chain, finding that two local
controls are still not sufficient for full controllability in the
XX-coupling case.  However, we can still get pure-state controllability
given two controls if the two controls are $Z_1$ and $X_2$ and the chain
length is $N=4k+2$, or $4k+3$.  In addition, we have shown that the
minimal control resource for full controllability is three local
controls acting on the first two spins.  This analysis provides a better
understanding of the relationship between Hamiltonian symmetries and the
system controllability, and minimal resources required to achieve
certain levels of controllability, which is important for both quantum
control theory as well as practical control of spin chains.
Furthermore, the techniques developed to compute the dynamical Lie
algebras explicitly by decomposing the Lie algebra into $n$-body
interaction terms, which can be iteratively generated from the original
Hamiltonians, will be useful to investigate controllability for other
spin networks where full controllability cannot be inferred from the
propagation property.

\section*{Acknowledgments}

We sincerely thank Dingyu Yang from New York University for his
contribution to the proof of Theorem~\ref{thm:rank}.  SGS acknowledges
funding from EPSRC ARF Grant EP/D07192X/1 and Hitachi. Part of this work
was carried out while DB held the EPSRC grant EP/F043678/1 at Imperial
College.

\appendix

\subsection{\label{A} $XXZ$ and $XYZ$ Chain with $Z_1$ Control}

We calculate the operators in $\L$ generated by $H_0$ and $H_1$ in
(\ref{eqn:xxz}):
\begin{align*}
[Z_1,H_0]&\rightarrow X_1Y_2-Y_1X_2\\
[Z_1,X_1Y_2-Y_1X_2]&\rightarrow X_1X_2+Y_1Y_2\\
[X_1X_2+Y_1Y_2,X_1Y_2-Y_1X_2]&\rightarrow Z_2-Z_1 \rightarrow Z_2\\
\cdots &\cdots
\end{align*}

Continuing the process, we can generate all $Z_k$, $X_kX_{k+1}+Y_kY_{k+1}$, and
$P(Z)=\sum_{j}^N \lambda_j\kappa_jZ_jZ_{j+1}$. Noticing that whenever we can generate
$X_kX_{k+1}+Y_kY_{k+1}$ we can also generate its conjugate term $X_kY_{k+1}-Y_kX_{k+1}$,
for simplicity, in the following we will only give the $X_kX_{k+1}+Y_kY_{k+1}$ terms.
Next, we have
\begin{align*}
&[X_kX_{k+1}+Y_kY_{k+1},X_{k+1}Y_{k+2}-Y_{k+1}X_{k+2}]\\
\rightarrow &X_kZ_{k+1}X_{k+2}+Y_kZ_{k+1}Y_{k+2}\\
\rightarrow &X_kZ_{k+1}Y_{k+2}-Y_kZ_{k+1}X_{k+2}\\
&[X_1Y_2-Y_1X_2,P(Z)]\rightarrow (X_1X_2+Y_1Y_2)Z_3\\
& [X_1Y_2-Y_1X_2,(X_1X_2+Y_1Y_2)Z_3]\rightarrow (Z_1-Z_2)Z_3\\
& [X_2Y_3-Y_2X_3,(Z_1-Z_2)Z_3]\rightarrow Z_1(X_2X_3+Y_2Y_3)\\
&[X_2Y_3-Y_2X_3,P(Z)]\\
\rightarrow &Z_1(X_2X_3+Y_2Y_3)+(X_2X_3+Y_2Y_3)Z_4\\
\rightarrow &(X_2X_3+Y_2Y_3)Z_4\\
&[X_2Y_3-Y_2X_3,(X_2X_3+Y_2Y_3)Z_4]\rightarrow (Z_2-Z_3)Z_4
\end{align*}
Continuing this process, we have 
\begin{align*}
&[X_3Y_4-Y_3X_4,(Z_2-Z_3)Z_4]\rightarrow Z_2(X_3X_4+Y_3Y_4)\\
&[X_3Y_4-Y_3X_4,H_2]\\
\rightarrow
&Z_2(X_3X_4+Y_3Y_4)+(X_3X_4+Y_3Y_4)Z_5\\
\rightarrow &(X_3X_4+Y_3Y_4)Z_5\rightarrow
(Z_3-Z_4)Z_5\\
&\cdots\cdots\cdots
\end{align*}
Thus, we can generate 
\begin{align*}
&Z_{k-1}(X_kX_{k+1}+Y_kY_{k+1})\\
&(X_kX_{k+1}+Y_kY_{k+1})Z_{k+2}\\
&Z_{k-1}(Z_k-Z_{k+1})\\
&(Z_k-Z_{k+1})Z_{k+2}
\end{align*}
Then we have 
\begin{align*}
&[Z_{k-1}(X_kX_{k+1}+Y_kY_{k+1}),X_{k-1}Y_k-Y_{k-1}X_k]\\
\rightarrow &(X_{k-1}X_{k+1}+Y_{k-1}Y_{k+1})\\
&[X_{k-1}X_k+Y_{k-1}Y_k, X_kY_{k+1}-Y_kX_{k+1}]\\
\rightarrow &(X_{k-1}X_{k+1}+Y_{k-1}Y_{k+1})Z_k\\
&[(X_{k-1}X_{k+1}+Y_{k-1}Y_{k+1})Z_k,(X_{k-1}Y_{k+1}-Y_{k-1}X_{k+1})]\\
\rightarrow &(Z_{k-1}-Z_{k+1})Z_k=Z_{k-1}Z_k- Z_kZ_{k+1},
\end{align*}
and together with $P(Z)$, we can decouple and generate all
$Z_kZ_{k+1}$. Moreover, we have
\begin{align*}
&[X_kY_{k+1}-Y_kX_{k+1},Z_{k+1}Z_{k+2}]\\ \rightarrow
&X_kX_{k+1}Z_{k+2}+Y_kY_{k+1}Z_{k+2}\\
&[X_kX_{k+1}Z_{k+2}+Y_kY_{k+1}Z_{k+2},X_{k+1}Y_{k+2}-Y_{k+1}X_{k+2}]\\
\rightarrow &X_kX_{k+2}+Y_kY_{k+2}\\
&[X_kY_{k+2}-Y_kX_{k+2},Z_{k+2}Z_{k+3}]\\ \rightarrow
&X_kX_{k+2}Z_{k+3}+Y_kY_{k+2}Z_{k+3}\rightarrow X_kX_{k+3}+Y_kY_{k+3}
\end{align*}
Continuing this processing, we can generate all $X_jX_k+Y_jY_k$. Furthmore, from
$X_jX_k+Y_jY_k$ and $X_mX_k+Y_mY_k$ we can generate $(X_jX_m+Y_jY_m)Z_k$. Hence we can
generate all $(Z_j-Z_m)Z_k$, and then any decoupled term $Z_mZ_k$.

Thus we can find all $k$-body operators: first, $Z_k$ is the only 1-body operator
we can generate. Then we can generate $X_jX_k+Y_jY_k$ and $X_jY_k-Y_jX_k$ as the
2-body operates. Next we can generate the 3-body XYZ-mixed operators
$(X_jX_m+Y_jY_m)Z_k$, from which we the 2-body $Z$ operator $(Z_j-Z_m)Z_k$. Then
based on the existing operators, we can generate the following 4-body XYZ-mixed
operators and 3-body $Z$ operators: 
\begin{align*}
&[(X_jX_k+Y_jY_k)Z_m,X_mY_n-Y_mX_n]\\
\rightarrow &(X_jX_k+Y_jY_k)(X_mX_n+Y_mY_n)\\
&[(X_jX_k+Y_jY_k)Z_m,Z_kZ_n]\rightarrow (X_jY_k-Y_jX_k)Z_mZ_n\\
&[X_jX_k+Y_jY_k,(X_jY_k-Y_jX_k)Z_mZ_n]\rightarrow (Z_j-Z_k)Z_mZ_n\\
\end{align*}

Continuing such process, we can sequentially generate $\ell$-body XYZ-mixed operators
and $(\ell-1)$-body $Z$ operators. When $\ell$ is even, we can generate: 
\begin{align*}
&(X_{m_1}X_{m_2}+Y_{m_1}Y_{m_2})Z_{m_3}\cdots Z_{m_\ell}\\
&(X_{m_1}X_{m_2}+Y_{m_1}Y_{m_2})(X_{m_3}X_{m_4}+Y_{m_3}Y_{m_4})Z_{m_5}\cdots
Z_{m_\ell}\\
& \cdots\cdots\cdots\\
&(X_{m_1}X_{m_2}+Y_{m_1}Y_{m_2})\cdots (X_{m_\ell-1}X_{m_\ell}+Y_{m_\ell-1}Y_{m_\ell})\\
&(Z_{m_1}-Z_{m_2})Z_{m_3}\cdots Z_{m_\ell}
\end{align*}
When $\ell$ is odd, we can generate:
\begin{align*}
&(X_{m_1}X_{m_2}+Y_{m_1}Y_{m_2})Z_{m_3}\cdots Z_{m_\ell}\\
&(X_{m_1}X_{m_2}+Y_{m_1}Y_{m_2})(X_{m_3}X_{m_4}+Y_{m_3}Y_{m_4})Z_{m_5}\cdots
Z_{m_\ell}\\
& \cdots\cdots\cdots\\
&(X_{m_1}X_{m_2}+Y_{m_1}Y_{m_2})\cdots
(X_{m_\ell-2}X_{m_\ell-1}+Y_{m_\ell-2}Y_{m_\ell-1})Z_{m_\ell}\\
&(Z_{m_1}-Z_{m_2})Z_{m_3}\cdots Z_{m_\ell}
\end{align*}

\subsection{\label{B} Rank of $M_{N,p}$}

\begin{theorem}\label{thm:rank}
\label{theorem:rank} For given $N$ and $p$ with $N\ge 2p$, the rank of
the $p$-pair operator set $E_{N,p}$ is equal to
$\binom{N}{p}\binom{N-p}{p}$.
\end{theorem}

Since, when evaluating $\rank(E_{N,p})$, only linear relations between
the operators of $E_{N,p}$ are involved, we can consider every element
of $E_{N,p}$ as a polynomial in terms of $2N$ variables, and transform
the original problem into evaluating the rank of a set of
polynomials. Specifically, for positive integers $N$ and $p$ with $N\ge
2p$, let $E$ be the set of any polynomials in terms of $2N$ variables
$x_1,x_2,\ldots,x_N$ and $y_1,y_2,\ldots, y_N$, satisfying the following
form:
\begin{align*}
&P[x_1,x_2,\ldots,x_N;y_1,y_2,\ldots, y_N]\\
=&q(m_1,m_2)q(m_3,m_4)\cdots
q(m_{2p-1},m_{2p})q(m_{2p-1},m_{2p})
\end{align*}
where $q(j,k)$ can take two forms, either $q(j,k)=x_jx_k+y_jy_k$ or
$q(j,k)=x_jy_k-y_jy_k$, and $m_k$'s are distinct from each other, with
$m_k\in [1,\ldots, N]$, $k=1,\ldots,2p$. In other words, any element in
$E$ is a product of $p$ terms, each taking the form $xx+yy$ or
$xy-yx$. For example, the following polynomials are in $E$:
\begin{align*}
(x_1x_2+y_1y_2)(x_3x_4+y_3y_4)\cdots(x_{2p-1}x_{2p}+y_{2p-1}y_{2p})\\
(x_1y_2-y_1x_2)(x_3y_4-y_3x_4)\cdots(x_{2p-1}y_{2p}-y_{2p-1}x_{2p})\\
\end{align*}
As discussed earlier, the total number of polynomials in $E$ is
$p!\binom{N}{p}\binom{ N-p}{p}$. However, not all of them are linearly
independent, and we aim to evaluate $\rank(E)$ over $\R$.

Since the $2N$ number of variables $x_j$ and $y_j$ are linearly
independent, the rank of $E$ over $\R$ is the same its rank over $\C$,
i.e., $\rank_{\R}(E)= \rank_{\C}(E)$. Next, over the field $\C$, we can
apply the following reversible transformations: $z_j = x_j+iy_j$ and
$z_j^*=x_j-iy_j$, and then all elements in $E$ can be expressed as
polynomials over $\C$ in terms of $z_j$ and $z_k^*$. This is equivalent
to considering raising/lowering operators in the algebra. Specifically,
\begin{align*}
x_jx_k+y_jy_k&=\Re(z_j^*z_k)\\
x_jy_k-y_jx_k&=\Im(z_j^*z_k)
\end{align*}
Then any element $q(m_1,m_2)\cdots q(m_{2p-1},m_{2p})$ in $E$ can be
rewritten as $Q(z_{m_1}^*z_{m_2})\cdots Q(z_{m_{2p-1}}^*z_{m_{2p}})$,
where $Q(z)$ is an operation that takes either the real or the imaginary
part of $z$.  Hence, all elements in $E$ can be rewritten in terms of
$2N$ number of independent complex variables: $z_1,\ldots,z_N$ and
$z_1^*,\ldots, z_N^*$. Next, we can show that the space generated by the
set $E$ is the same as the one generated by the set $F$ whose elements
are in the following form: $z_{m_1}^*z_{m_2}\cdots
z_{m_{2p-1}}^*z_{m_{2p}}$, where $m_k\in [1,\ldots, N]$,
$k=1,\ldots,2p$. In order to see this, we will show that
$Q(z_{m_1}^*z_{m_2})\cdots Q(z_{m_{2p-1}}^*z_{m_{2p}})$ in $E$ can be
generated by the elements in $F$: from the sum of or the difference
between $z_{m_1}^*z_{m_2}\cdots z_{m_{2p-1}}^*z_{m_{2p}}$ and
$z_{m_2}^*z_{m_1}\cdots z_{m_{2p-1}}^*z_{m_{2p}}$, we can generate
$Q(z_{m_1}^*z_{m_2})z_{m_3}^*z_{m_4}\cdots
z_{m_{2p-1}}^*z_{m_{2p}}$. Continuing such process, we can use
polynomials in the form of $Q(z_{m_1}^*z_{m_2}) z_{m_3}^*z_{m_4}\cdots
z_{m_{2p-1}}^*z_{m_{2p}}$ to generate elements in the form of
$Q(z_{m_1}^*z_{m_2}) Q(z_{m_3}^*z_{m_4})\cdots
z_{m_{2p-1}}^*z_{m_{2p}}$. Continuing such process we can finally
generate $Q(z_{m_1}^*z_{m_2}) \cdots Q(z_{m_{2p-1}}^*z_{m_{2p}})$. On
the other hand, reversing such process, we can generate
$z_{m_1}^*z_{m_2}\cdots z_{m_{2p-1}}^*z_{m_{2p}}$ from
$Q(z_{m_1}^*z_{m_2})\cdots Q(z_{m_{2p-1}}^*z_{m_{2p}})$. Therefore, we
have shown that $\Span(E)=\Span(F)$ over $\C$, inducing
$\rank_{\C}(E)=\rank_{\C}(F)$.

Next, we evaluate the rank of $F$ over $\C$. Since the $2N$ variables
$z_j$ and $z_j^*$ are linearly independent, all elements in $F$, as in
the product form of $z_{m_1}^*z_{m_2}\cdots z_{m_{2p-1}}^*z_{m_{2p}}$,
are hence independent as well. In order to obtain an element in $F$, we
choose $p$ number of $z_j$'s from the indices $j\in \{1,\ldots,N\}$ and
choose $p$ number of $z_j^*$'s from the remaining $N-p$ number of
indices. Then the number of elements in $F$ is
$\binom{N}{p}\binom{N-p}{p}$, which equals to
$\rank_{\C}(F)=\rank_{\C}(E)=\rank_{\R}(E)$. Thus, we have proved
$\rank_{\R}(E)\binom{N}{p}\binom{N-p}{p}$, and hence the result of
Theorem~\ref{theorem:rank} follows.

\subsection{\label{C} Dimension of $\L$}

After we derive all the $k$-body operators in $\L$, we evaluate
$\dim(\L)$, where the following identity is involved:
\begin{lemma}\label{combina}
\begin{align*}
\binom{2N}{N}=\sum_{p=0}^{\fl{N/2}} \frac{N!}{p!^2(N-2p)!}2^{N-2p}
\end{align*}
\end{lemma}

\begin{IEEEproof}
Define a polynomial $f(x)=(x+1)^{2N}$. Then the term $x^N$ has coeffient $\binom{2N}{N}$.
We have 
\begin{align*}
f(x)&=(x^2+2x+1)^2\\
&=\sum_{p+q+r=N}\frac{N!}{p!q!r!}(x^2)^p(2x)^q\\
&=\sum_{0\le p+q\le N}\frac{N!}{p!q!(N-p-q)!}x^{2p+q}2^q
\end{align*}
Hence,
\begin{align*}
\binom{2N}{N}&=\sum_{0\le p+q\le N,2p+q=N}\frac{N!}{p!q!(N-p-q)!}2^q\\
&=\sum_{0\le p+q\le N,2p+q=N}\frac{N!}{p!(N-2p)!p!}2^{N-2p}\\
&=\sum_{p=0}^{\fl{N/2}}\frac{N!}{p!^2(N-2p)!}2^{N-2p}
\end{align*}
\end{IEEEproof}


\begin{thebibliography}{09}

\bibitem{Burgarth-Bose}
D.~Burgarth, S.~Bose, C.~Bruder, and V.~Giovannetti, Phys.~Rev.~A \textbf{79}, 060305(R) (2009).

\bibitem{Burgarth-10}
R. Heule, C. Bruder, D. Burgarth and V. M. Stojanovic, Phys.~Rev.~A \textbf{82}, 052333 (2010) (2010).

\bibitem{Alastair}
A.~Kay and P.~J.~Pemberton-Ross, Phys.~Rev.~A \textbf{81}, 010301 (2010).

\bibitem{Burgarth}
D.~Burgarth, \emph{et al.}, Phys. Rev. A \textbf{81}, 040303(R)
(2010).

\bibitem{Thomas}
U. Sander and T.~Schulte-Herbr\"uggen, J. Math. Phys. 52, 113510 (2011).

\bibitem{WS_symmetry}
X. Wang and S. G. Schirmer, to appear in IEEE Trans. Autom. Control, 2012, arXiv:1012.3695 (2010).

\bibitem{D'Alessandro-book}
D. D'Alessandro, \emph{Introduction to Quantum Control and Dynamics}
(Chapman \& Hall/CRC, Boca Raton, 2008)

\bibitem{uni_qc}
M. Nielsen and I. Chuang, \emph{Quantum Computation and Quantum Information} 
(Cambridge University Press, 2000)

\bibitem{sussmann}
H.~J.~Sussmann and V.~Jurdjevic, J. Diff. Equations. \textbf{12} , 95 (1972).

\bibitem{WSBB}
X. Wang, A. Bayat, S. G. Schirmer and S. Bose, Phys. Rev. A \textbf{81}, 032312 (2010).

\bibitem{Jacobson}
N. Jacobson, \emph{Lie Algebras} (John Wiley, New York, 1962)

\bibitem{pure_control} 
F. Albertini and D. D'Alessandro, Notions of controllability for multi-level quantum-mechanical systems, 
IEEE Trans. Autom. Control \textbf{48}, 1399 (2003);
%Proc. 40th IEEE Conference on Decision and Control, \textbf{2} 1589 (2001) and 
S. G. Schirmer, A. I. Solomon and J. V. Leahy, J. Phys. A, \textbf{35}, 4125 (2002).

\end{thebibliography}
\end{document}